\documentclass{emulateapj}

\shorttitle{$r$-Process in proto-neutron-star wind}

\shortauthors{Wanajo}

\begin{document}

\title{The \lowercase{$r$}-process in proto-neutron-star wind revisited}

\author{Shinya Wanajo\altaffilmark{1}
        }

\altaffiltext{1}{National Astronomical Observatory of Japan,
        2-21-1 Osawa, Mitaka, Tokyo 181-8588, Japan;
        shinya.wanajo@nao.ac.jp}

\begin{abstract}
We examine the $r$-process in the neutrino-driven proto-neutron-star
(PNS) wind of core-collapse supernovae in light of the recent findings
of massive neutron stars in binaries as well as of an indication of
neutron-richness in the PNS ejecta because of the nucleon potential
corrections on neutrino opacities. To this end, a spherically symmetric,
general relativistic, steady-state wind model is applied for a wide
range of PNS masses between $1.2 M_\odot$ and $2.4 M_\odot$ with the
latter reaching the causality limit. Nucleosynthesis calculations with
these PNS models are performed by assuming a time evolution of electron
fraction with its minimal value of $Y_\mathrm{e} = 0.4$, which mimics
recent hydrodynamical results. The fundamental nucleosynthetic aspect of
the PNS wind is found to be the production of Sr, Y and Zr in
quasi-equilibrium and of the elements with $A \approx 90$--110 by a weak
$r$-process, which can be an explanation for the abundance signatures in
$r$-process-poor Galactic halo stars. PNSs more massive than $2.0
M_\odot$ can eject heavy $r$-process elements, however, with
substantially smaller amount than what is needed to account for the
solar content. PNS winds can be thus the major origin of light
trans-iron elements but no more than 10\% of those heavier than $A \sim
110$, although they may be the sources of the low-level abundances of Sr
and Ba found in numerous metal-poor stars if the maximum mass of PNSs
exceeds $2.0 M_\odot$.

\end{abstract}


\section{Introduction}

Proto-neutron-star (PNS) wind of core-collapse supernovae (CCSNe), the
outflows driven by neutrino heating, has long been suggested to be the
major site of the $r$-process (rapid neutron-capture process) since
early 1990's \citep{Meyer1992, Woosley1994}. One of the problems in
these early works was the very high entropy in the wind, $S \sim 400\,
k_\mathrm{B}\, \mathrm{nucleon}^{-1}$ ($k_\mathrm{B}$ is the Boltzmann
constant), which was not confirmed by subsequent works \citep[$S \sim
100\, k_\mathrm{B}\, \mathrm{nucleon}^{-1}$,][]{Takahashi1994,
Qian1996}. General relativistic effects were found to increase entropy
\citep{Cardal1997} but needed a PNS more massive than $M \approx 2.0
M_\odot$ for robust $r$-processing \citep{Otsuki2000, Wanajo2001,
Thompson2001}. Another problem was an unacceptable overproduction of
some species such as Sr, Y, and Zr \citep{Woosley1994, Wanajo2001}. More
seriously, hydrodynamical simulations of CCSNe with elaborate neutrino
transport indicated proton-richness in the wind ejecta
\citep{Fischer2010, Huedpohl2010}. These works seemed to exclude the PNS
wind scenario as the $r$-process site.

Recent works on the effect of nucleon potential corrections for neutrino
opacities seem, in part, to revive the PNS wind scenario
\citep{Reddy1998, Roberts2012a, Martinez2012, Roberts2012b,
Horowitz2012}. These works predict that the electron fraction
($Y_\mathrm{e}$; number of protons per nucleon) drops off from an
initially proton-rich value to the minimal value of $\sim 0.42$--0.45
and increases again towards $\sim 0.5$ in the late wind phase, the
behavior not expected in previous works. Recent discoveries of massive
NSs in binary systems with a precision measurement of $M = 1.97 \pm 0.04
M_\odot$ for PSR~J1614-2230 \citep{Demorest2010} and an inferred mass of
$M \sim 2.4 M_\odot$ for PSR~B1957+20 \citep{vanKerkwijk2011} are also
encouraging for the PNS wind scenario.

In this Letter we aim to revisit the issue of the $r$-process in PNS
winds in light of these recent findings. This is to extend and improve
the previous nucleosynthesis studies, which were based on limited
hydrodynamical outcomes \citep[e.g.,][]{Woosley1994, Arcones2011} or
semi-analytic solutions with a few selected parameter sets \citep[e.g.,
$M = 1.4 M_\odot$ and $2.0 M_\odot$;][]{Wanajo2001} with $Y_\mathrm{e}$
evolutions that were incompatible with the recent works. To this end, a
semi-analytical wind model \citep{Wanajo2001} is applied for a wide
range of $M$ between $1.2 M_\odot$ and $2.4 M_\odot$
(Section~2). Nucleosynthesis calculations are performed with the wind
solutions by assuming a time evolution of $Y_\mathrm{e}$ (Section~3)
that mimics the result of \citet{Roberts2012b}. The nucleosynthesis
yields are mass-integrated to compare with the solar $r$-process
abundances as well as those in Galactic halo stars. We then discuss
whether PNS winds can be the sources of $r$-process elements in the
Galaxy.

\section{Wind model}\label{sec:model}

\begin{figure*}
\epsscale{1.0}
%\epsscale{0.8}
%\plotone{property.eps}
\plotone{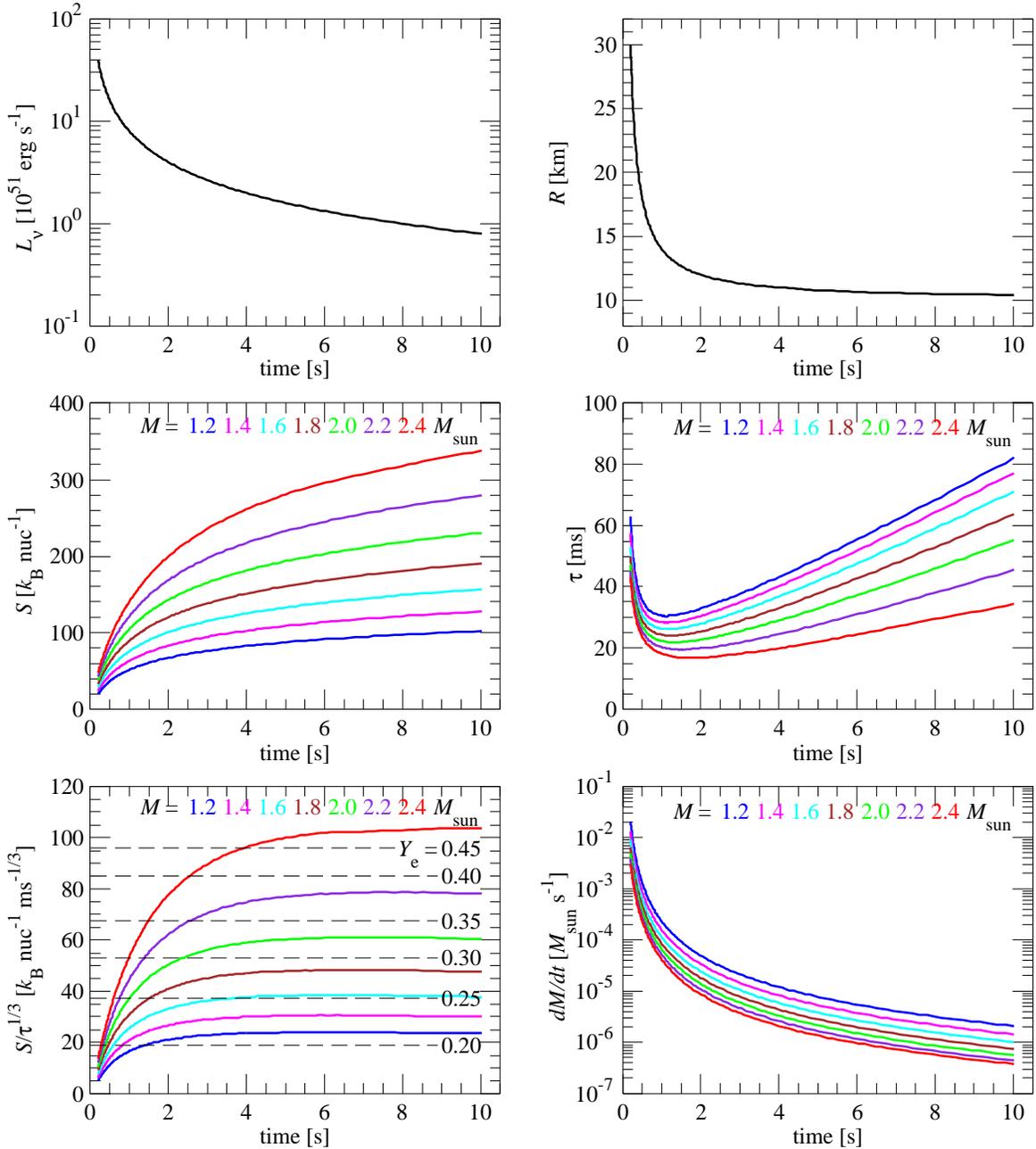}
\caption{Time evolutions of $L_\nu$ (top left) and $R$ (top right)
 adopted in this study. Resulting $S$ (middle left), $\tau$ (middle
 right), $S/\tau^{1/3}$ (bottom left), and $\dot{M}$ (bottom right) are
 shown as functions of $t$. In the
 bottom-left panel, the $Y_\mathrm{e}$'s, above which the production of
 $A \sim 200$ nuclei are predicted, are indicated by dashed lines.
}
\label{fig:property}
\end{figure*}

\begin{figure}
\epsscale{1.0}
%\epsscale{0.8}
%\plotone{yeg.eps}
\plotone{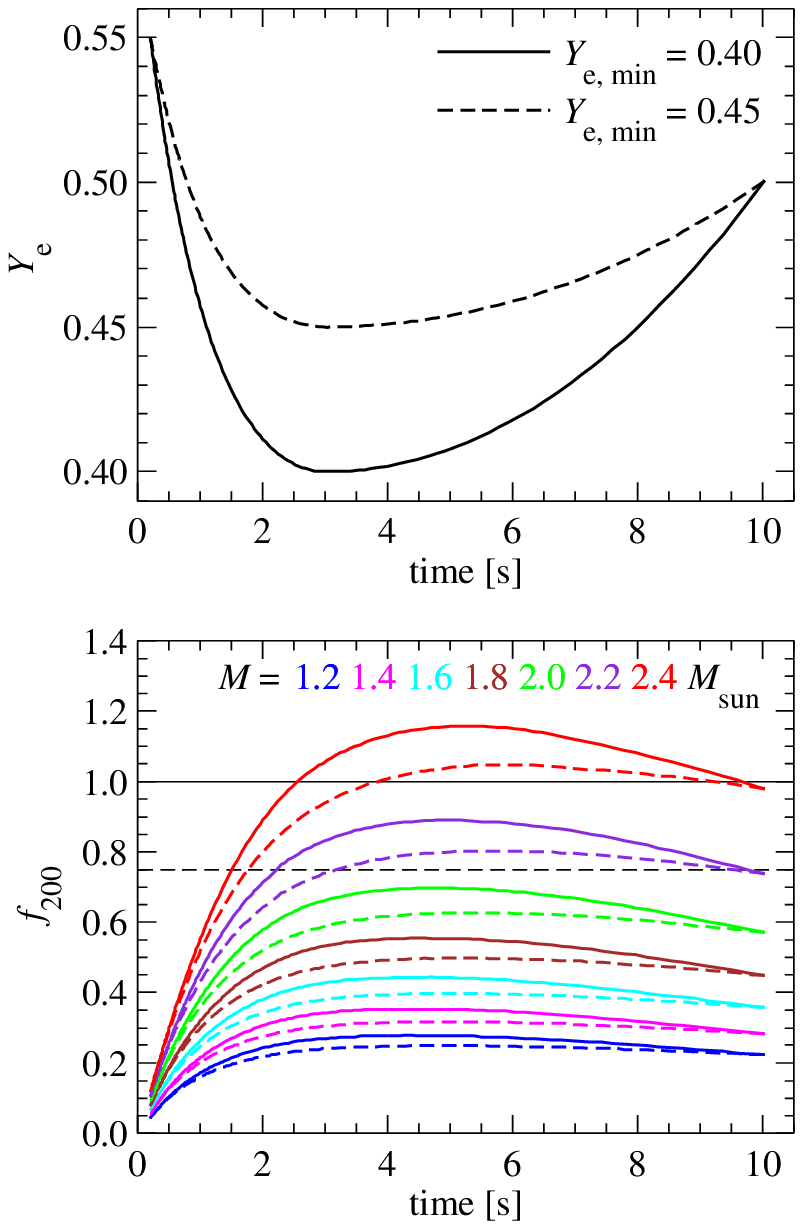}
\caption{Top: time evolution of $Y_\mathrm{e}$ adopted for nucleosynthesis calculations
 ($Y_\mathrm{e, min} = 0.40$; solid curve). The case with $Y_\mathrm{e,
 min} = 0.45$ is also shown by the dashed curve. Bottom: values of
 $f_{200}$ (Eq.~[1]) as functions of $t$.
 The cases for $Y_\mathrm{e, min} = 0.40$ and 0.45 are shown by
 solid and dashed curves, respectively. The solid and
 dashed horizontal lines indicate $f_{200} = 1$ and $f_{130} = 1$,
 respectively, above
 which the production of $A \sim 200$ and $A \sim 130$ nuclei are expected.
}
\label{fig:yeg}
\end{figure}

Although the underlying physics what causes the explosions of CCSNe has
been under debate \citep[e.g.,][]{Janka2012}, it is known that the
neutrino-driven outflows after evacuation of the early convective ejecta
are well described by the steady-state (semi-) analytical solutions of
PNS wind models \citep{Duncan1986, Qian1996, Cardal1997, Otsuki2000,
Wanajo2001, Thompson2001}. In this study, we use the spherically
symmetric, general relativistic, semi-analytic wind model in
\citet[][for more detail, see their Section~2]{Wanajo2001}.
% The neutrino
% heating ($\nu_e$ and $\bar{\nu}_e$ captures on free nucleons, neutrino
% scattering by $e^-$ and $e^+$, and $\nu\bar{\nu}$ pair annihilation to
% $e^-e^+$ pairs) and cooling ($e^-$ and $e^+$ captures on free nucleons
% and $e^-e^+$ pair annihilation to $\nu\bar{\nu}$ pairs) rates with full
% considerations of general relativistic effects (i.e., the redshift of
% neutrino energies and bending of the neutrino trajectories) are adopted
% from \citet{Otsuki2000}. 
The average neutrino energies are taken to be 12, 14, and 14~MeV for
electron neutrino, electron antineutrino, and heavy lepton neutrinos,
respectively \citep[according to, e.g.,][]{Fischer2010,
Huedpohl2010}. The equation of state for ions (ideal gas) and
arbitrarily degenerate, arbitrarily relativistic electrons and positrons
is taken from \citet{Timmes2000}.

Each wind (i.e., transonic) solution can be obtained for a given set of
$(M, R, L_\nu)$; we assume the PNS radius $R$ to be the same as the
neutrinosphere and the neutrino luminosities of all the flavors to have
the same value $L_\nu$. We consider the models of $M/M_\odot = 1.2$,
1.4, 1.6, 1.8, 2.0, 2.2, and 2.4, which cover the range of measured and
estimated masses of NSs \citep{Demorest2010, vanKerkwijk2011,
Lattimer2011}. For $L_\nu$ and $R$, phenomenological time evolutions
during the first 10~s after core bounce ($t_0 = 0.20\, \mathrm{s} \le t
\le t_1 = 10\, \mathrm{s}$) are adopted as follows. To roughly mimic
recent results of long-term simulations over the PNS cooling phase
\citep[e.g.,][]{Fischer2010, Huedpohl2010}, we assume $L_\nu (t) =
L_{\nu, 0} (t/t_0)^{-1}$ with $L_{\nu, 0}\, (\mathrm{erg~s}^{-1}) =
10^{52.4} = 3.98 \times 10^{52}$ (Fig.~\ref{fig:property}; top left). We
also assume $R(L_\nu) = (R_0 - R_1) (L_\nu/L_{\nu, 0}) + R_1$ with $R_0
= 30$~km and $R_1 = 10$~km so that each wind solution can be obtained
from a given set of $(M, L_\nu)$. This is equivalent to set $R(t) = (R_0
- R_1) (t/t_0)^{-1} + R_1$ (Fig.~\ref{fig:property}; top right). Note
that $R(t_1) = 10.4$~km matches the lower bound of the constraint for
cold NSs (with $M = 1.4 M_\odot$), $10.4\, \mathrm{km} \le R \le 12.9\,
\mathrm{km}$, inferred by \citet{Steiner2013}. With this assumption, the
$2.4 M_\odot$ model reaches the smallest radius ($\approx 10.3$~km)
allowed by the causality limit (the speed of sound must not exceed the
speed of light), $R \gtrsim 4.3\, (M/M_\odot)$~km
\citep{Lattimer2011}. The PNS with $M = 2.4 M_\odot$ should be thus
taken as the absolute extreme model.\footnote{\citet{Wanajo2001} showed
that, with inclusion of general relativistic effects, nucleosynthetic
outcomes were roughly scaled with $M/R$. Nucleosynthetic results of
$M/M_\odot = 1.2$--2.4 with $R_1 = 10$~km here would be thus similar to
those of, e.g., $M/M_\odot = 1.4$--2.9 with $R_1 = 12$~km.}

Wind solutions for each $M$ model are computed for $\log\, L_\nu
(\mathrm{erg~s}^{-1}) = 52.60, 52.59, \dots, 50.90$ ($t_0 \le t \le
t_1$; 171 $L_\nu$'s). The middle and bottom panels
(Fig.~\ref{fig:property}) illustrate the resulting basic properties. We
confirm the previous results \citep{Otsuki2000, Wanajo2001,
Thompson2001} that the asymptotic entropy ($S$; middle left) increases
with time, being systematically greater for more massive PNSs. We
further find a strong sensitivity of $S$ to $M$ for $> 2.0 M_\odot$,
which reaches $338\, k_\mathrm{B}\, \mathrm{nucleon}^{-1}$ for $M = 2.4
M_\odot$. This is a consequence of the general relativistic effects that
are particularly important when $M/R$ is close to the causality limit
\citep{Cardal1997}. We also find systematically smaller expansion
timescales \citep[$\tau$; defined as the $e$-folding time of temperature
below 0.5~MeV,][]{Otsuki2000} for more massive PNSs, which take minimal
values at $t \sim 1$--2~s and increase with time
(Fig.~\ref{fig:property}; middle right).

As analytically shown by \citet{Hoffman1997}, the ratio $S/\tau^{1/3}$
(with a fixed $Y_\mathrm{e}$) serves as the measure of the strength of
$r$-processing. We find in the bottom-left panel
(Fig.~\ref{fig:property}) that $S/\tau^{1/3}$ increases with time and
saturates at $t \sim 5$~s as a result of the increasing $\tau$ that
counterbalances the increasing $S$. That is, the strength of
$r$-processing becomes mostly independent of time for $t \gtrsim 5$~s if
$Y_\mathrm{e}$ is kept constant. The dashed lines indicate the values of
$Y_\mathrm{e}$ above which the production of $A \sim 200$ nuclei are
expected, according to the analytical formula in \citet[][their
Eq.~(19)]{Hoffman1997}. We find that an unacceptably low $Y_\mathrm{e}$
($< 0.25$) is required for $M = 1.4 M_\odot$. If a currently predicted
minimal value of $Y_\mathrm{e} \sim 0.42$--0.45 were taken, one would
need a PNS with the mass close to the extreme case of $M = 2.4 M_\odot$
for the production of heavy $r$-process
nuclei.\footnote{\citet{Otsuki2000} and \citet{Wanajo2001} showed that
the wind of $(M, L_\nu) = (2.0 M_\odot, 10^{52} \mathrm{erg~s}^{-1})$
with $Y_\mathrm{e} = 0.40$ led to a robust $r$-process. Their PNS radius
was, however, fixed to 10~km, which was appreciably smaller than our
more reasonable, time-dependent value of $R(L_\nu = 10^{52}
\mathrm{erg~s}^{-1}) = 15$~km.}  The bottom-right panel shows the mass
ejection rates ($\dot{M}$) that are systematically smaller for more
massive PNS models and quickly decrease with time. This indicates that
the wind ejecta are dominated by the early components with small
$S/\tau^{1/3}$. The very late ejecta for $t > 10$~s (if any; not
considered in this study) would be unimportant.

The time evolution of $Y_\mathrm{e}$, which is needed for
nucleosynthesis calculations, is assumed as $Y_\mathrm{e}(t) = c_1
\cosh[c_2(t-t_\mathrm{min})] + c_3$ (Fig.~\ref{fig:yeg}; top), where
$c_2 = 1.0$ for $t < t_\mathrm{min} = 3.0$~s and $c_2 = 0.10$ for $t >
t_\mathrm{min}$. The coefficients $c_1$ and $c_3$ are determined to
satisfy $Y_\mathrm{e}(t_\mathrm{min}) = Y_\mathrm{e, min}$ and, for $t <
t_\mathrm{min}$ and $t > t_\mathrm{min}$, respectively,
$Y_\mathrm{e}(t_0) = 0.55$ and $Y_\mathrm{e}(t_1) = 0.50$. This roughly
mimics the hydrodynamical result by \citet[][see their
Fig.~5]{Roberts2012b}. We adopt $Y_\mathrm{e, min} = 0.40$ (solid curve
in the top panel of Fig.~\ref{fig:yeg}), the value slightly smaller than
$\sim 0.42$--0.45 in \citet{Roberts2012b}. While $Y_\mathrm{e}$ drops in
response to the PNS contraction, the increasing neutrinospheric density
suppresses the charged current neutrino interactions by Pauli blocking
and $Y_\mathrm{e}$ cannot decrease at late times
\citep{Fischer2012}. Note that the $\alpha$-effect
\citep{McLaughlin1996, Meyer1998}, which were not considered in
\citet{Roberts2012b}, would slightly shift $Y_\mathrm{e}$ towards $\sim
0.5$. The value of $Y_\mathrm{e, min}$ here may thus be taken as the
absolute lower limit for PNS winds.

The lower panel (Fig.~\ref{fig:yeg}) shows the condition for making the
third peak nuclei ($A \sim 200$) according to \citet{Hoffman1997},
\begin{eqnarray}
f_{200} = 
\frac{(S / 230\, k_\mathrm{B}\, \mathrm{nucleon}^{-1})}
{(Y_\mathrm{e}/0.40)(\tau / 20\, \mathrm{ms})^{1/3}} \gtrsim 1,\,
0.38 \lesssim Y_\mathrm{e} \lesssim 0.46.
\end{eqnarray}
This reflects the value of $Y_\mathrm{e}$ in addition to the combination
$S/\tau^{1/3}$ (Fig.~\ref{fig:property}; bottom left). We find that only
the extreme model of $M = 2.4 M_\odot$ satisfies this condition (the
region above the horizontal solid line). Also indicated by the
horizontal dashed line is the condition for making the second peak ($A
\sim 130$) nuclei, $f_{130} \approx 1.34\, f_{200} \gtrsim 1$. It
indicates that only the models with $M \gtrsim 2.0 M_\odot$ can reach
the second peak of the $r$-process abundances. The $f_{200}$ curves with
$Y_\mathrm{e, min}$ replaced by 0.45 are also shown in
Figure~\ref{fig:yeg}, implying slightly weaker $r$-processing.

\section{Nucleosynthesis}\label{sec:nuc}

\begin{deluxetable}{cccccccc}
\tablecaption{Ejecta masses (in units of $10^{-5} M_\odot$)}
\tablewidth{0pt}
\tablehead{
\colhead{$M/M_\odot$} &
\colhead{1.2} &
\colhead{1.4} &
\colhead{1.6} &
\colhead{1.8} &
\colhead{2.0} &
\colhead{2.2} &
\colhead{2.4}
}
\startdata
total     & 219  & 143  & 100  & 74.1  & 56.7    & 44.6    & 36.0    \\
$^{4}$He  & 122  & 92.7 & 71.9 & 56.9  & 45.8    & 37.4    & 31.0    \\
%$^{64}$Zn &     0.485852 &     0.214457 &     0.091203 &     0.036573 &
% 0.013226 &     0.004660 &     0.001733 \\
%$^{92}$Mo &     0.000170 &     0.000104 &     0.000059 &     0.000030 &
% 0.000014 &     0.000005 &     0.000002 \\
$A > 100$ & 2.19 & 2.75 & 2.76 & 2.27  & 1.78    & 1.37    & 0.893   \\
Sr        & 3.61 & 1.92 & 1.09 & 0.627 & 0.346   & 0.177   & 0.0764  \\
Ba        & 0.00 & 0.00 & 0.00 & 0.00  & 0.0420  & 0.0373  & 0.0199  \\
Eu        & 0.00 & 0.00 & 0.00 & 0.00  & 0.00452 & 0.00585 & 0.00305
\enddata
\end{deluxetable}

\begin{figure}
\epsscale{1.0}
%\epsscale{0.8}
%\plotone{ymav.eps}
\plotone{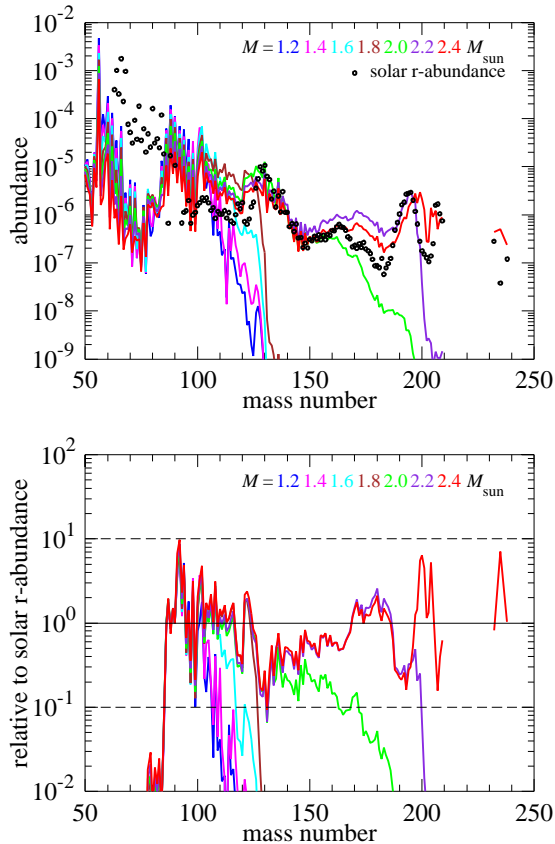}
\caption{Top: mass-integrated nuclear abundances, which are compared with
 the solar $r$-process abundances (circles) that shifted to match the
 third peak height ($A \sim 200$) for the $2.4 M_\odot$ model. Bottom:
 ratios of mass-integrated abundances relative to the solar $r$-process
 abundances (scaled at $A = 90$).
}
\label{fig:ymav}
\end{figure}

\begin{figure*}
\epsscale{1.0}
%\epsscale{0.8}
%\plotone{weakr.eps}
\plotone{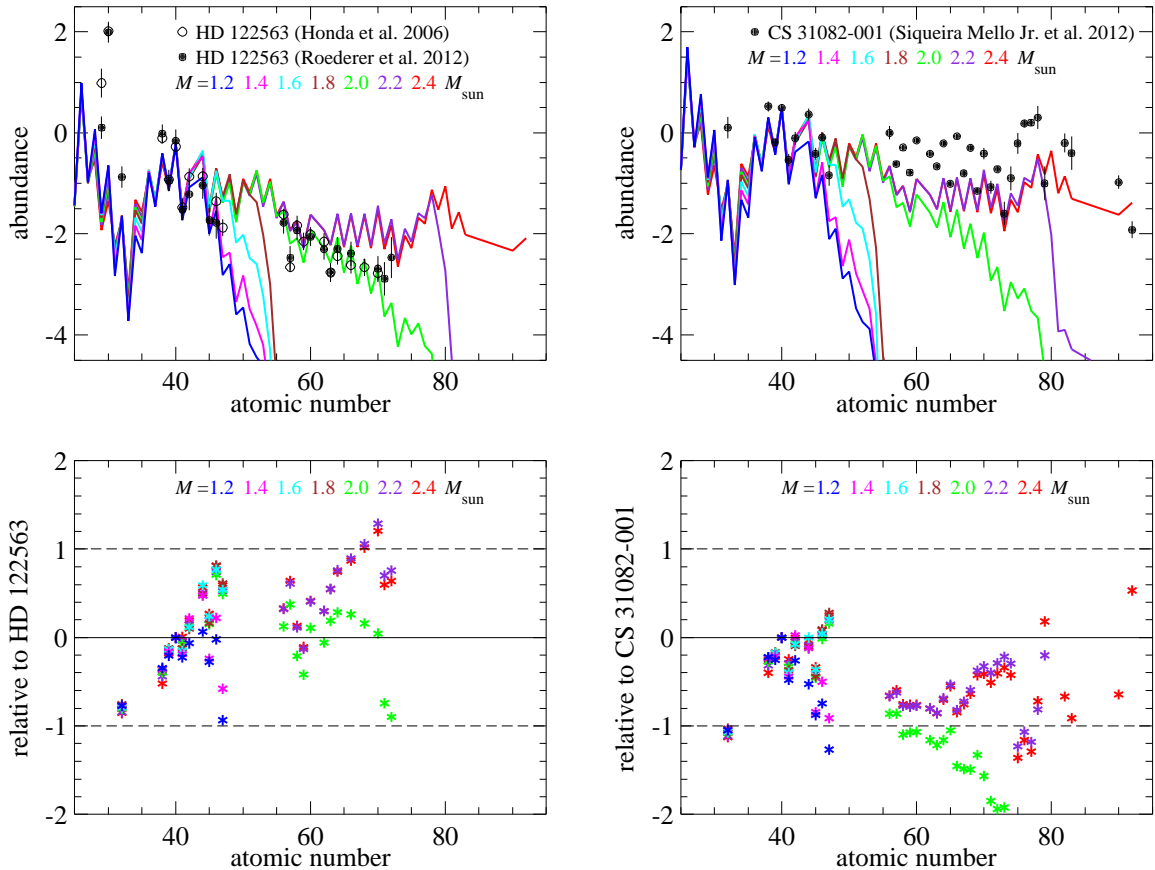}
\caption{Mass-integrated elemental abundances compared to the 
 stellar abundances (top panels) and their ratios (bottom panels). Two Galactic
 halo stars, HD~122563 \citep[left,][]{Honda2006, Roederer2012} and
 CS~31082-001 \citep[right][]{Siqueira2013} are taken as
 representative of $r$-process-poor/rich stars, respectively. The
 nucleosynthetic abundances are normalized at $Z = 40$.
}
\label{fig:weakr}
\end{figure*}

The nucleosynthetic yields for all the $(M, L_\nu)$ sets are computed
with the reaction network code described in \citet{Wanajo2001,
Wanajo2011b}. Reaction rates are employed from the latest library of
REACLIB V2.0 \citep{Cyburt2010} for the experimental evaluations when
available and the rest from the theoretical estimates in BRUSLIB
\citep{Xu2013} based on the HFB-21 mass predictions
\citep{Goriely2010}. The $\beta$-decay rates are taken from the gross
theory predictions \citep[GT2][]{Tachibana1990} obtained with the HFB-21
masses. Neutrino interactions, which would slightly shift $Y_\mathrm{e}$
by the $\alpha$-effect,
% but play
%sub-dominant roles for nucleosynthesis in PNS winds
are not included. Using thermodynamic trajectories of PNS winds, the
calculations are started when the temperature decreases to 10~GK,
assuming initially free protons and neutrons with mass fractions
$Y_\mathrm{e}$ and $1 - Y_\mathrm{e}$, respectively.\footnote{We
examined only the $Y_\mathrm{e, min} = 0.40$ case with $t_\mathrm{min} =
3.0$~s. Tests showed that a small shift of $t_\mathrm{min}$ did not
qualitatively change our result. The cases with $Y_\mathrm{e, min} =
0.45$ corresponded to roughly re-scaling $M$'s with $\sim 0.1$--$0.2
M_\odot$ smaller values. Note also that the presence of the preceding SN
ejecta that give rise to the termination-shocks \citep{Arcones2007} do
not change the gross abundance features \citep{Wanajo2007, Kuroda2008}.}

The nucleosynthetic abundances are mass-integrated (Fig.~\ref{fig:ymav};
top) by adopting $\dot{M}$ for each PNS model. For comparison purposes,
the solar $r$-process compositions (circles) are also plotted to match
the third peak height ($A \sim 195$) for the $M = 2.4 M_\odot$ model. As
anticipated from the lower panel of Figure~\ref{fig:yeg}, only the
extreme model of $M = 2.4 M_\odot$ satisfactorily accounts for the
production of heavy $r$-process nuclei up to Th ($A = 232$) and U ($A =
235$ and 238). The $2.2 M_\odot$ model reaches the third peak abundances
but those beyond. The $2.0 M_\odot$ model reaches the second ($A \sim
130$) but the third peak abundances. We find no strong $r$-processing
for the models with $M < 2.0 M_\odot$.

We find, however, quite robust abundance patterns below $A \sim 110$,
which appears a fundamental aspect of nucleosynthesis in PNS winds. The
double peaks at $A \approx 56$ and 90 with a trough between them are
formed in quasi-nuclear equilibrium (QSE; $\gtrsim 4$~GK). Note also
that the overproduction of $N = 50$ species $^{88}$Sr, $^{89}$Y,
$^{90}$Zr \citep{Woosley1994, Wanajo2001} are not prominent in our
result. This is due to the short duration of moderate $S$ ($< 100\,
k_\mathrm{B}\, \mathrm{nucleon}^{-1}$; Fig.~\ref{fig:property}) with
$Y_\mathrm{e} \sim 0.45$ (Fig.~\ref{fig:ymav}), in which the $N = 50$
species copiously form in QSE. The lower panel of Figure~\ref{fig:ymav}
shows the ratios of nucleosynthetic abundances relative to their solar
$r$-process values (normalized at $A = 90$). For $2.2 M_\odot$ and $2.4
M_\odot$ models, the ratios are more or less flat between $A = 90$ and
200, although deviations from unity are seen everywhere.
%which could be, in part, due to the incompleteness in astrophysical
%modelings here as well as in nuclear data for neutron-rich nuclei. 
%This is a consequence of the fact that the early winds are assumed to be
%proton-rich and then $Y_\mathrm{e}$ drops off to the minimal value in a
%short period of time (Fig.~\ref{fig:ymav}; top). 

Table~1 provides the masses (in units of $10^{-5} M_\odot$) of the total
ejecta, $^4$He, those with $A > 100$, Sr, Ba, and Eu, for all the PNS
models. The total ejecta masses span a factor of 6 with smaller values
for more massive PNSs. The larger fractions of $^4$He in more massive
models, however, lead to the ejecta masses for $A > 100$ (total masses
of $r$-process nuclei) ranging only a factor of 2.5. The masses of Sr
range a factor of 50 with the greater amount for less massive models. Ba
and Eu are produced only in the massive models with $M \ge 2.0 M_\odot$.

Studies of Galactic chemical evolution estimate the average mass of Eu
per CCSN event (if they were the origin) to be $\sim 10^{-7} M_\odot$
\citep{Ishimaru1999}, that is, $\sim \mathrm{a~few} 10^{-5} M_\odot$ for
the nuclei with $A > 100$. Taken at the face value, the Eu masses for $M
\ge 2.0 M_\odot$ reach 30\%--60\% of this requirement. The fraction of
events with such massive PNSs would be limited to no more than $\sim
20\%$ of all CCSN events (e.g., $\gtrsim 25 M_\odot$). The masses of Eu
from these massive PNSs are, therefore, about 10 times smaller than the
requirement from Galactic chemical evolution (the same holds for
Ba). Note that, for massive PNS cases, the ejecta masses would be
further reduced by fallback or black-hole formation \citep{Qian1998,
Boyd2012}. For Sr, the required mass per CCSN event is estimated to be
$\sim 2 \times 10^{-6} M_\odot$ from the solar $r$-process ratio of
Sr/Eu = 16.4 \citep{Sneden2008}. The low mass PNS models, which may
represent the majority of CCSNe, thus overproduce Sr by about a factor
of 10. The amount of QSE products such as Sr, Y, and Zr is, however,
highly dependent on the multi-dimensional $Y_\mathrm{e}$ distribution in
early times \citep[$t < 1$~s,][]{Wanajo2011a}.

Figure~\ref{fig:weakr} compares the mass-integrated abundances with
those of Galactic halo stars.  
% Early works have
% reveiled that all the studied $r$-process-rish stars have the
% ``universal'' abundance patterns that resemble that of the solar
% $r$-process abundances \citep[at least for those heavier than
% Ba,][]{Sneden2008}. Subsequent works have shown, however, that
% $r$-process-poor stars, that are in fact the majority of extremely
% metal-poor stars, have quite different patterns from the solar
% $r$-process curve \citep{Honda2006}. 
Two well known objects are taken as representative of $r$-process-poor
\citep[HD~122563, left panels;][]{Honda2006, Roederer2012} and
$r$-process-rich \citep[CS~31082-001, right panels;][]{Siqueira2013}
stars with the metallicities [Fe/H] $= -2.7$ and $-2.9$,
respectively. These stars have [Eu/Fe] $= -0.52$ and $+1.69$,
respectively, well below and above the average value of $\approx +0.5$
at [Fe/H] $\approx -3$. The top and bottom panels show, respectively,
the mass-integrated abundances and their ratios relative to the stellar
abundances, which are normalized to the stellar abundances at $Z = 40$.

In the left panels, we find that the $1.2 M_\odot$ and $1.4 M_\odot$
models result in reasonable agreement with the stellar abundances
between $Z = 38$ (Sr) and $Z = 48$ (Cd). The $2.0 M_\odot$ model nicely
reproduces the abundance pattern of HD~122563 up to $Z = 68$ (Er) but
somewhat overproduces the elements of $Z = 46$--48 (Pd, Ag, Cd). It
could be thus possible to interpret that the abundance signatures of
$r$-process-poor stars were due to a weak $r$-process that reaches $Z
\sim 50$ ($M < 2.0 M_\odot$) or 70 ($M = 2.0 M_\odot$) with or without
additional sources for $Z > 50$, respectively. In the right panels, we
find that the stellar abundances between $Z = 38$ (Sr) and $Z = 47$ (Ag)
are well reproduced by massive models with $M \ge 1.6 M_\odot$. The
models with $M = 2.2 M_\odot$ and $2.4 M_\odot$ produce the heavier
elements with a similar pattern to that of CS~31082-001 but with a
smaller ratio. Because of the insufficient production of Eu (Table~1),
our PNS models would not account for the high [Eu/Fe] value in this
star. The winds from such massive PNSs ($M \gtrsim 2.0 M_\odot$) could
be, however, still the source of the low-level abundances (factor of
several 10 smaller than the average values) of Sr and Ba in numerous
metal-poor stars \citep{Roederer2013, Aoki2013}.

\section{Conclusion}

We revisited the issue of the $r$-process in neutrino-driven PNS winds
in light of recent findings of the massive NSs as well as of the
neutron-richness in the PNS ejecta. Nucleosynthesis calculations were
performed with the semi-analytical wind models over a wide range of the
PNS masses ($1.2 \le M/M_\odot \le 2.4$), assuming a phenomenological
time evolutions of $Y_\mathrm{e}$.

Based on our result, including the extreme model ($M = 2.4 M_\odot$)
that encounters the causality limit, it would be safe to conclude that
neutrino-driven PNS winds were excluded as the major origin of heavy
$r$-process elements. Note that some previous works have suggested
several mechanisms that help to increase $S$ and reduce $\tau$, e.g.,
strong magnetic field \citep{Thompson2003, Suzuki2005} and highly
anisotropic neutrino emission \citep{Wanajo2006}. Our conclusion would
not be changed if such mechanisms (associated to probably rare events)
were considered, as far as the dominant driving source of the wind is
neutrino heating (which sets $\dot{M}$). Other driving mechanisms such
as magnetorotationally \citep{Metzger2007, Winteler2012} or
acoustic-wave driven outflows are beyond the scope of this Letter and
cannot be excluded as the $r$-process origins on the basis of our
result.

Neutrino-driven PNS winds are, however, promising sources that eject
light trans-iron elements made in QSE (Sr, Y, and Zr) and by a weak
$r$-process (up to Pd, Ag, and Cd), in addition to the early convective
ejecta of CCSNe \citep{Wanajo2011a}. If not the main origin of the
$r$-process elements beyond $A \sim 110$, they could be the sources of
low-level abundances of Sr and Ba (e.g., [(Sr, Ba)/Fe] $< -1$) found in
extremely metal-poor stars. Future surveys or indications of massive
NSs, long-term hydrodynamical simulations of PNS winds, and galactic
chemical evolution studies will be important to make clear the role of
PNS winds in the enrichment histories of galaxies.

This work was supported by the JSPS Grants-in-Aid for Scientific
Research (23224004).

\end{document}